\documentstyle[aps,prb,twocolumn]{revtex}

\begin{document}

\bibliographystyle{prsty}


\wideabs{

\title{Lessons from vortex dynamics in super media}

\author{ P. Ao }
\address{ Departments of Mechanical Engineering and Physics, 
          University of Washington, Seattle, WA 98195, USA }
\date{ June 30, 2004 }

\maketitle

\begin{abstract}
 Some aspects of vortex dynamics theories are critically examined. The discussion is placed in the context of experiments on the Josephson-Anderson effect and on the Hall anomaly in the mixed state. 
\end{abstract}


}


\section{Introduction}

One of all time beautiful examples of physical theories motivated by logical consistency is Dirac`s theory of relativistic quantum electrodynamics: The void in the filled Fermi sea led Dirac to predict a new particle, later called positron. The influence of its success is now on all branches of modern physics.  Another beautiful example is Abrikosov`s vortex lattice theory, inferred as a very natural consequence, in the hindsight, of the Ginzburg-Landau theory.  Beautiful theories can also be obtained at the opposite end, by careful consideration and critical analysis of experimental data. One of the all time such examples is the Bardeen-Cooper-Schrieffer theory of superconductivity. Again, its influence is now on all branches of modern physics. 

Daily theoretical physicists` activities are usually between those two extremes.
For example, by combining limited experimental data with theoretical insights, difficult aspects of the vortex dynamics puzzle were solved by Nozieres and Vinen \cite{nv} and by Bardeen and Stephen \cite{bs} in the 1960`s.  Those works are such elegant illustrations that all researchers interested in vortex dynamics field should study them carefully.  

As we marvel at those achievements, uneasiness has been accumulated. 
While the vortex friction in the Bardeen-Stephen work appears to be largely consistent with experimental data, the transverse force in the Nozieres-Vinen work seems not.  
To be precise, if viewing vortex dynamics in real superconductors as the same of {\bf  independent} vortices, that is, no correlation among vortices,  the Nozieres-Vinen work would predict a large Hall angle. Experimentally, however, the Hall angle is usually small. More puzzling is that the Hall angle can change sign, sometimes even more than once.   In the early 1970`s, it became clear that those experimental data were real, not experimental errors. 
This pronounced and apparent discrepancy between theory and experiment is hence the famous Hall anomaly in the mixed state of superconductors. 
Similar discrepancies have been observed in neutral superfluids.

In order to resolve this Hall anomaly, even after it had been shown by Ao and Thouless, and their coworkers \cite{at,az}that based on the microscopic theory and global topological analysis the total transverse force on an individual vortex is indeed large, the large transverse force in the Nozieres-Vinen work has been questioned. Subsequently it has been concluded by a large group of eminent theorists that the total transverse force on a single moving vortex must be usually small and occasionally changes sign. Various independent vortex dynamics theories have been developed during the past 30 years. Such efforts have been conveniently summarized in two recent reviews \cite{v1,v2}. 
It is clear, nevertheless, that despite those ingenious efforts to fit experimental data, the Hall anomaly still remains an `anomaly` in the light of those independent vortex dynamics theories.  What goes wrong?

\section{Four Mathematical Inconsistencies}

A careful examination of those theories reveals that there are two types of errors in those theories in Ref`s.[\onlinecite{v1,v2}]. The first type lies in physics: The vortex many-body correlations are completely absent in their explanations of Hall anomaly by independent vortex dynamics theories.  
The hint to correct this independent particle dynamics error is in fact already implied in the work of Dirac and Abrikosov but absent in those independent vortex dynamics theories, to which I will come back later.  The second type of errors lies in mathematics: 
In order to fit experimental data on Hall anomaly, the mathematical consistency in those theories has been severely compromised. This ignorance of mathematical consistency goes also against the spirit of Dirac and Abrikosov. Let me explain the mathematical inconsistencies in those independent vortex dynamics theories first.

A further close analysis of those theories suggests that the mathematical inconsistencies in those independent vortex dynamics theories may be classified into four different categories. 

\subsection{Misuse of the relaxation time approximation}

The most subtle inconsistency is the relaxation time approximation employed by Kopnin and his co-workers (Kopnin and Kratsov, JETP Lett.,1976; Kopnin and Lopatin, Phys. Rev. B, 1995; van Otterlo, Feigelman, Geshkenbein, Blatter, Phys. Rev. Lett., 1995; documented in Ref.[6]).
This type of mistakes frequently occurs in the force-balance type calculation of transport coefficients, which was already noticed at least by Green in the 1940`s and has been extensively discussed by Kubo. 
Unfortunately, as Kubo remarked in his coauthored famous book on statistical physics, in the literature such error repeatedly appears in different disguises.  

It would be very attempting to use such a simple relaxation time approximation following the typical diagrammatic technique. But, it is plainly wrong in the present context of microscopic derivation of vortex dynamics based on force balance equation. 
Fortunately, a careful calculation of  the vortex friction and its transverse force with disorders ranging from weak to strong was performed by Ao and Zhu \cite{az}. 
No relaxation time approximation is needed. The disorder effect can be directly taken into account.
The Nozieres-Vinen and Bardeen-Stephen theories have been mathematically united in this work \cite{az}. It is also a detailed implementation of the global topological method \cite{at}. 
One may now stop the use of erroneous relaxation time approximation in the derivation of vortex dynamics.
 
\subsection{Mixing with another effect}

The second type of mistake is very interesting. It has been claimed that there is a contribution from the normal fluid which would cancel (or add to, depending on the interpretation) the large transverse force determined by the superfluid density (Sonin Soviet Phys. JETP,1976; Phys. Rev. B, 1997, documented in Ref.[5]). From the very beginning questions have been raised on the interpretation of the result and on the validity of various approximations. For the clarity of the present discussion let us accept the view that the normal fluid can indeed be represented by phonons and that phonons scattering off a vortex can indeed be exactly mapped onto the Aharonov-Bohm scattering problem.  

It has been known at least since Aharonov and Bohm that such a scattering is periodic in magnetic flux. When transforming this result back into vortex dynamics, this means that the transverse force according to such phonon contributions is a periodic function of vorticity, completely different from that of the transverse force represented by the Magnus force \cite{nv,at,az} which is linear in vorticity.    
Hence the calculation of such phonon contributions is of a completely different kind. 
It corresponds to a different experimental condition.  In fact, if we should conceive the similar situation of a vortex scattering of a superfluid island, a periodic dependence on the enclosed superfluid particle numbers has already been obtained by Ao and coworkers \cite{zta} as well as by many others. 

In the derivation of vortex dynamics by geometric phase computation, the phase vortex acquired is a {\bf continuous distribution} of the vortex trajectory, and the phase in the situation conceived by Sonin would be a {\bf discontinuous} function. 
This difference explains that in the former case the transverse force is a linear function of the superfluid density while in the latter it is not.  Those two situations should not be mixed. This difference is a fine manifestation of symmetry breaking for solutions of a symmetric Hamiltonian.

\subsection{Double counting the same topological effect}

The third mistake is the claim of the spectral flow canceling the Berry phase (Volovik, JETP Lett., 1993 and1997, documented in both Ref.[5] and [6]).  This mistake could be easily committed by anyone if not being careful and critical enough:  There are two ways of calculating the topological contribution for the transverse force in fermionic superfluids: One is far from the vortex core. It is the method employed by Ao and Thouless and further refined by Thouless, Ao, and Niu \cite{at}. Another is from the core, the spectrum flow, also related to the curvature or connection. They are equivalent according to Stokes theorem \cite{ao97}.  Hence they should not be used as different forces to cancel each other.  
As is now well established, the Magnus force is a manifestation of the Josephson-Anderson relation. It is worth mentioning that it has already been well known that for fermionic superfluids phase slippage of the Josephson relation is equivalent to the spectrum flow.

It is perhaps also useful to point out that in modern physics there are several significant topological phenomena which can be computed in seemingly completely different ways.
A fine example is the transport in quantum Hall effect in condensed matter physics. There it is firmly established that the calculation of Hall conductance due to edge states is equivalent to one due to bulk states. In one particular experimental situation one type of calculation may be more straightforward. Nevertheless, no one would like to have those two calculations cancel each other in Hall conductance. 
%
%

The mistake of the spectral flow canceling the Berry phase is a double counting of the same effect.

\subsection{No extra Berry phase contribution at vortex core}

The fourth mistake is rather mysterious, perhaps associated with the difficulty in the understanding of the Anderson theorem in dirty superconductors.
It  was claimed (Feigelman, Geshkenbein, Larkin, Vinokur, JETP Lett, 1995, documented in Ref.[6] ) that  there were two contributions to Berry phase, one far from the core, the same as that of Ao and Thouless, and an extra one from the vortex core, because the corresponding superfluid density would be finite at the point of phase singularity. 
The starting point to demystify this mistake is quantum mechanics: 
At the phase singularity the wavefunction must be zero, hence the superfluid density at the core of the corresponding phase singularity must be zero. 
There is simply no extra Berry phase from the core because the corresponding superfluid density is zero. 

Let us analyze this mistake from two different levels of description: macro- and micro-scopic.  On the level of the phenomenological Ginzburg-Landau type description, which is now believed to be valid for both clean and dirty superconductors, the superfluid density is obviously zero at the vortex core. Therefore, there is no extra contribution to the Berry phase. Now, a person familiar with Bardeen-Cooper-Schrieffer microscopic theory would say that this macroscopic description is an oversimplication. The superfluid density is finite at the vortex core even for a super super clean superconductor.  Indeed, the superfluid density is finite at the vortex core according to the Bardeen-Cooper-Schrieffer microscopic theory. The same microscopic theory also tells us that first, for this finite superfluid density at the vortex core there is no associated phase singularity, and, second, if there is a phase singularity associated with a quasiparticle state, its wavefunction is zero at this singularity, as would be expected from quantum mechanics. Hence, whether or not a finite density exists, there is no extra Berry phase from the vortex core. 
This is true for both super clean and dirty superconductors. 

\section{Magnus Force and Josephson-Anderson Relation}

We have now explained that the microscopically derived independent vortex dynamics theories documented in Ref.[5] and [6] are all mathematically inconsistent. Even if the reader agrees with this strong conclusion, one important question still remains: the experimental evidence. 

There is actually already a substantial body of experimental supports for the works of Nozieres-Vinen and Bardeen and Stephen, as well as that by Ao and Thouless, and their coworkers. I use the word `substantial` because there is a big group of eminent physicists, such as discussed above, would not accept those experiments.  Also, it is needless to say that further experiments are needed. 
Given this consideration, I mention here three important experiments. The first one is the classical experiment to establish the quantization of vorticity, assuming that the total transverse force is indeed proportional to the superfluid density which can be measured by different means \cite{vinen}. This experiment was published in 1961, and is a Josephson type experiment before Josephson`s predictions.     
The second type of experiment is a systematic measurement of the effect of a moving vortex, the Magnus force and its closely related Josephson(-Anderson) relation \cite{packard}.
The third one is a direct measurement of transverse force of a moving vortex in a superconductor \cite{zhu}. All those experiments clearly establish the large transverse force on a moving vortex.

\section{Hall Anomaly and Vortex Many Body Effect}

However, one may insist that the Hall anomaly still remains unexplained:  
How could the large transverse force lead to a small Hall angle, and, sometime to a sign change?  The ingredients to explain the Hall anomaly are actually already implied in the works of Dirac and Abrikosov.  

Let us stand one step away from vortex dynamics and consider the Hall effect in semiconductors.  The transverse force there, the Lorentz force, on a moving electron is universal. There exist such extremely rich Hall phenomena in semiconductors: small Hall angle, sign change, Quantum Hall effect, etc. 
It would be puzzling that how could a universal Lorentz force generate such a complexity.  Fortunately, this puzzle has long been solved: The competition between electron many body effect (Coulomb interaction, Fermi-Dirac statistics) and pinning (lattice, impurities, etc).  
The key to solve the puzzle is a logical extension of Dirac`s idea of a void in a filled Fermi sea: the existence of holes in a filled energy band \cite{mermin}.

The ubiquitous existence of the Abrikosov lattice, the starting point of theoretical considerations in the mixed state of superconductors, already loudly suggests that one must consider vortex many body effect.   Following this suggestion, it has been argued that the competition between vortex many body interaction and pinning can explain the Hall anomaly in the mixed state \cite{ao95}. Starting from this idea, it is rather straightforward to work out quantitative predictions on the Hall anomaly, even with the simplest types of collective excitations in an Abrikosov lattice such as vacancies and interstitials:
The scaling laws, the activation energies, the sign change, etc. More complicated many body effects are clearly possible to play a role. This vortex many body effect consideration is also consistent with the tremendous progress in the last 10 years in the explanation of the longitudinal resistance by vortex many body effect. This successful explanation of Hall anomaly clearly implies that the independent vortex dynamics model is not the only model for Hall anomaly and that the use of it to explain Hall anomaly is physically inconsistent with both the Abrikosov lattice theory and the recent theoretical progress in vortex matter. We may further point out that the vortex many body effect-pinning model appears consistent with all major experimental observations on Hall anomaly. But, it has not been universally accepted yet. As a good sign, it turns out that two of the strong advocators of independent vortex dynamics theories already believe the vortex many body effect can indeed lead to the sign change \cite{kv}.

\section{A Lesson}

Thus, the vortex dynamics theory started by Nozieres and Vinen, by Bardeen and Stephen, and further developed by Ao and Thouless, and their coworkers, works well. Four types of technical mistakes in those independent vortex dynamics summarized in Ref.[5] and [6] aiming to fit Hall anomaly data have been discussed above. 
For a reader not in the immediately related fields, the technical issues above are not particularly helpful. What other lessons can one learn?  

A generic lesson can indeed be learned. As discussed at the beginning, there are two opposite methods in theoretical physics research: emphasizing the logical consistency or emphasizing the direct hints from experiments.
Both can and have lead to great successes.  However, pushing each method to its extreme can also be very problematic. For example, in pushing his unified field theory for elementary particles, Heisenberg claimed to have a final theory for physics and the only job left to be done would be to work out the details. We note that more than 40 years later string theorists are still busy and working hard on `details` of different final theories.  
I am afraid that the major mistake made by those eminent researchers in their independent vortex dynamics theories on Hall anomaly is another extreme: 
In pushing those theories to fit Hall anomaly data, the mathematical consistency of their theories is greatly compromised.  60 years later we are still working hard on the foundation of vortex dynamics \cite{ao03}. 

Is there a take-home rule on how to successfully use those two opposite methodologies?
The fun and challenge of practicing theoretical research, as well as its pitfall, are that there is no ready formula for the right mix of two extremes. You can only find it out by getting your hands wet and dirty, preferably at the spot where the water is the roughest \cite{weinberg}.



\begin{thebibliography}{99}
\bibitem{note}
   If a reader feels a need for a more detailed list of works discussed in Section II, 
I apologize for this inconvenience. The main rational for this omission is to keep the reference list short. It is also because there exists a complete coverage of those works in Ref`s.[\onlinecite{v1,v2}].
The rational to put a modest list of works by Ao and/or Thouless on the transverse force and on the Hall anomaly is that they are not referenced in Ref`s.[\onlinecite{v1,v2,kv}]. 
For a reader interested in their study, Ref.[\onlinecite{az}] may be a good starting point.
\end{thebibliography}
\end{document}